\documentclass[12pt]{iopart}

\usepackage{iopams}
\usepackage{float}
\usepackage{graphicx}

\newcommand{\NVm}{NV$^-$}

\begin{document}

\title[Probing of diamond NV ensembles coupled to GaP microcavities]{Low-temperature tapered-fiber probing of diamond NV ensembles coupled to GaP microcavities}

\author{K.-M.C. Fu$^{1,2}$, P.E. Barclay$^{3,4}$, C. Santori$^1$, A. Faraon$^1$ and R.G. Beausoleil$^1$}
\address{$^1$Information and Quantum System Lab, Hewlett-Packard
Laboratories, 1501 Page Mill Road, Palo Alto, CA 94304, USA}
\address{$^2$Department of Physics, University of Washington, Seattle, WA 98195, USA}
\address{$^3$Institute for Quantum Information Science and Department of Physics and Astronomy,
University of Calgary, Calgary, AB T2N 1N4, Canada }
\address{$^4$NRC-National Institute for Nanotechnology, Edmonton, AB T6G 2M9, Canada}

\ead{kaimeifu@uw.edu}

\begin{abstract}
In this work we present a platform for testing the device performance of a cavity-emitter system, using an ensemble of emitters and a tapered optical fiber. This method provides high-contrast spectra of the cavity modes, selective detection of emitters coupled to the cavity, and an estimate of the device performance in the single-emitter case. Using nitrogen-vacancy (NV) centers in diamond and a GaP optical microcavity, we are able to tune the cavity onto the NV resonance at 10~K,  couple the cavity-coupled emission to a tapered fiber, and measure the fiber-coupled NV spontaneous emission decay. Theoretically we show that the fiber-coupled average Purcell factor is 2-3 times greater than that of free-space collection; although due to ensemble averaging it is still a factor of 3 less than the Purcell factor of a single, ideally placed center.

\end{abstract}


%78.66.Fd Optical properties of specific thin films -> III-V semiconductors
%42.82.Et Integrated optics -> Waveguides, couplers, and arrays
%61.72.jn Defects and impurities in crystals: microstructure -> color centers
%03.67.Lx Quantum computation architectures and implementations
%33.50.Dq Fluorescence and phosphorescence spectra
%42.25.Kb Coherence
%42.50.Ar Photon statistics and coherence theory
%42.50.Ct Quantum description of interaction of light and matter; related experiments
%42.50.Dv Quantum state engineering and measurements
%42.50.Ex Optical implementations of quantum information processing and transfer in quantum optics
%42.50.Pq Cavity quantum electrodynamics; micromasers
%61.72.J- Point defects and defect clusters
%76.30.Mi Color centers and other defects
%78.67.-n Optical properties of low-dimensional, mesoscopic, and nanoscale materials and structures
%78.67.Hc Quantum dots

\maketitle

\section{Introduction}

Several properties of the negatively-charged nitrogen-vacancy center (\NVm) in diamond make it an attractive candidate as a qubit for quantum information processing (QIP). First, the ground-state electron-spin decoherence time exceeds a millisecond in isotopically purified $^{12}$C diamond samples~\cite{ref:balasubramanian2009usc}. Second, an excited-state manifold~\cite{ref:neumann2009ess,ref:tamarat2008sfs} provides optical access to both initialize and control the electron ground-state spin~\cite{ref:buckley2010slc, ref:santori2006cptb} as well as a way to create quantum entanglement through photon interference~\cite{ref:cabrillo1999ces}. Finally, hyperfine coupling of the electron spin to either N  or $^{13}$C nuclear spins~\cite{ref:neumann2008mea,ref:gurudevdutt2007qrb} provides access to a second storage qubit needed for quantum protocols based on repeat until success schemes~\cite{ref:childress2005ftq, ref:benjamin2006bgs}.

The majority of \NVm~experiments have been performed on NV centers found deep within the diamond substrate utilizing free-space excitation and collection optics. There is now significant motivation to integrate these quantum emitters into optical devices to build scalable measurement-based optical quantum networks. To realize these networks coupling \NVm~centers to optical cavities will be necessary to enhance emission into the zero-phonon line (ZPL)~\cite{ref:santori2010nqo} and for coupling ZPL emission into waveguides for efficient collection. NV-cavity coupling can be obtained by coupling NV centers in diamond nanoparticles to non-diamond microcavities~\cite{ref:wolters2010ezp, ref:barclay2009cie,ref:park2006cqd,ref:englund2010dcs}, coupling NV centers in single-crystal diamond to non-diamond cavities (hybrid approach)~\cite{ref:barclay2009cbm, ref:larsson2009com}, and by fabricating microcavities out of single-crystal diamond (all-diamond approach)~\cite{ref:faraon2010rez,ref:bayn2011tnp,ref:hiscocks2008dwf,ref:babinec2010dns}. Due to the poor spectral properties of NV centers found in nano-diamond, the first approach is challenging for scalable optical networks that rely on interference for `which-path' erasure. Of the single-crystal diamond approaches, the all-diamond approach has the advantage of a higher optical field strength at the NV site which lies within the cavity structure, and significant enhancement of the NV emission into the ZPL has already been observed in all-diamond microrings~\cite{ref:faraon2010rez}. The hybrid approach, which relies on coupling diamond NVs to the evanescent field extending out of the non-diamond cavity, has the potential advantage of additional device functionality provided by the material properties of the waveguide layer. For example, the electro-optic properties of the cavity material could enable switching or cavity tuning capabilities.

 A figure of merit for the coupling between a cavity mode and dipole emitter is the spontaneous emission enhancement (Purcell) factor $F_p$. For an emitter at position $\vec r$ in a monolithic cavity and on resonance with the cavity mode,
\begin{equation}
F_{p} = \frac{3}{4\pi^2}\frac{\lambda_0^3}{n_c^3}\frac{Q}{V}\left|\frac{E(\vec r)}{ \vec E (\vec r_\mathrm{max})}\right|^2,
\label{eq:Purcell_simple1}
\end{equation}
in which $\lambda_0^3$ is the vacuum wavelength, $n_c$ is the cavity refractive index, $Q$ is the cavity mode quality factor, $V$ is the cavity mode volume,  $|\vec E(\vec r)|$ is the electric field strength at the dipole,  and $|\vec E (\vec r_\mathrm{max})|$ is the maximum electric field strength. There are several key device factors that must be optimized. The optical cavity quality factor $Q$ must be sufficiently large and its mode volume $V$ sufficiently small. The quantum emitter must be coupled spatially to the cavity mode and spectrally on resonance. Finally, in addition to cavity-dipole coupling, the cavity mode must be efficiently coupled to a useful input/output mode (i.e. a waveguide).

When engineering optical devices for QIP applications, device testing becomes prohibitively difficult when all testing must be done on single emitters. This is especially true in the common case where the spectral and spatial properties of the emitter are difficult to control. In this work we develop a test platform based on emitter ensembles designed to evaluate coupled emitter-cavity performance. This platform is particularly useful for hybrid-type devices in which the material properties of the waveguide layer may need to be tested before investing the resources necessary to test single-emitter devices.

In this work we focus on coupling an ensemble of near-surface \NVm~centers in single crystal diamond to GaP micro-cavities fabricated through photolithographic and dry-etching techniques. We show that it is possible using a tapered fiber to collect emission preferentially from centers that are coupled to the cavity~\cite{ref:knight1997pme,ref:srinivasan2005pmq}. We are able to deterministically tune the cavity resonance several nanometers onto the ~\NVm~ZPL and perform lifetime measurements using the photoluminescence (PL) collected through the tapered fiber. The lifetime measurements can be used to evaluate NV-device coupling. Although we are experimentally unable to observe a significant reduction in the ensemble lifetime for these particular devices, we demonstrate the feasibility of the testing method and theoretically show that there is a significant advantage to using the tapered fiber over free-space collection in ensemble measurements.

\section{Hybrid GaP-diamond cavity fabrication}

\begin{figure}
 \centering
  \centerline{\scalebox{0.25}{\includegraphics{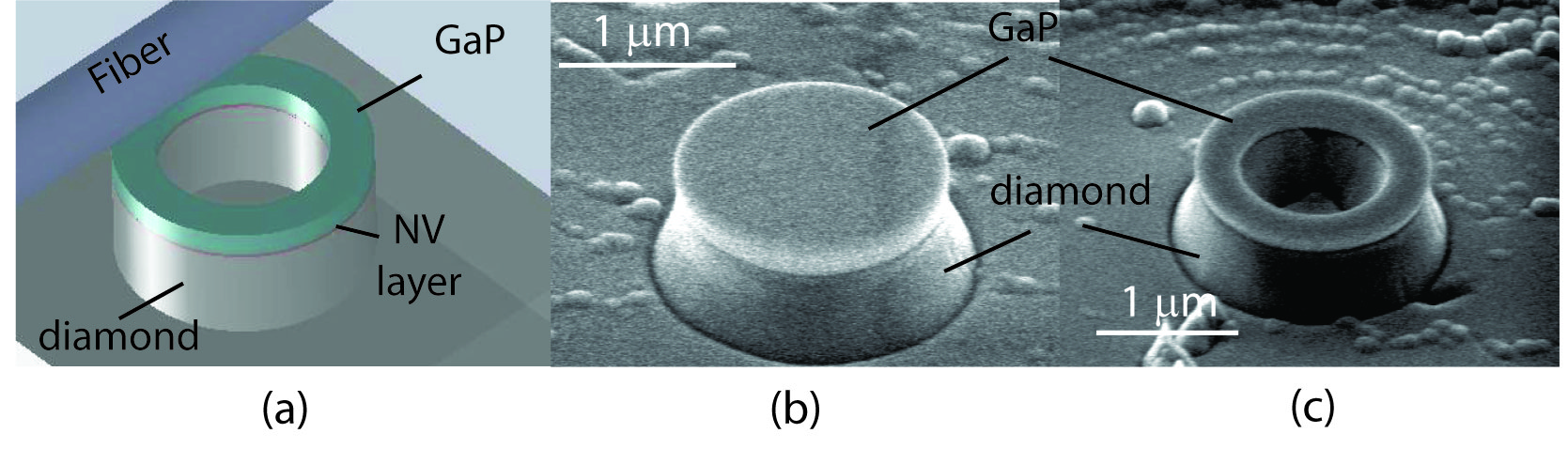}}}
  \caption{(a) Schematic of tapered-fiber coupling to GaP-diamond hybrid microcavity. (b) Device D1: 1.5~$\mu$m diameter GaP disk with thickness 250~nm. (c) Device D2: 1.6~$\mu$m outer diameter GaP ring with thickness 150~nm. }
  \label{fig:devices}
\end{figure}

The tested devices consisted of a GaP microcavity on top of a diamond pedestal with an ensemble of NVs near the diamond surface (Fig.~\ref{fig:devices}a). GaP has a refractive index of $n_\mathrm{GaP}=3.3$ that is higher than that of diamond ($n_\mathrm{D} = 2.4$), and it is transparent at 637nm, the ZPL wavelength. Thus devices made from GaP on diamond can support high-Q modes that couple evanescently to the NV centers near the diamond surface.

Experiments were performed on a a commercially available type IIa (110)-oriented diamond chip (Element 6) grown by chemical vapor deposition (CVD) with a nitrogen concentration of [N]$<$1~ppm as specified by the manufacturer. The NV ensemble was created by N$^+$ ion implantation at 10~keV (CORE Systems) at a dose of $2\times10^{13}$~cm$^{-2}$ followed by a one hour high temperature anneal at 900~$^\circ$C in a H$_2$/Ar forming gas. To maximize the fraction of NV centers in a negatively-charged state, the sample was further annealed in an oxygen atmosphere at 465~$^\circ$C for 2 hours~\cite{ref:fu2010cnn}. In this sample there are three mechanisms of NV formation. First, NVs are incorporated during the CVD growth process. The density of these NVs is expected to be less than $10^{14}$~cm$^{-3}$~\cite{ref:acosta2009dhd} and thus their contribution to the total collected signal can be neglected. A second way an NV is formed is when a vacancy created by the implantation process diffuses and is trapped by a native nitrogen during the annealing process. The maximum NV density due to this process is 2$\times10^{15}$~cm$^{-3}$ based on electron irradiation studies aimed at maximizing the \NVm~yield in similar CVD material~\cite{ref:acosta2009dhd}. These NVs are located within the first 200~nm of the surface~\cite{ref:santori2009vdn} and have a maximum sheet density of $\sim4\times10^{10}$~cm$^{-2}$. Finally, an NV can form when a vacancy is trapped by implanted nitrogen. This third type of NV is expected to be $14\pm5$~nm from the surface, as estimated using SRIM~\cite{ref:zeigler2008sri}. The estimated sheet density is $10^{12}$~cm$^{-2}$ assuming a 10\% conversion rate of implanted ions to \NVm~which is typical in our devices. Even for free-space collection the majority of the NV PL signal is due to the implanted nitrogen. By using a tapered fiber we are able to further select NVs which are coupled to the cavity. We note that if an electronic-grade CVD diamond substrate ([N]$<$1~ppb)  were used only the third type of NV would be present.

The starting material for the GaP microcavities was a commercially available (IQE) wafer consisting of a 250~nm GaP layer on an 800~nm Al$_{0.8}$Ga$_{0.2}$P sacrificial layer grown by molecular beam epitaxy. For one set of devices the original 250~nm thickness was used while in a second set the GaP layer was thinned by dry etching to 150~nm. The microcavities were defined photolithographically using a mask consisting of 1.5-3~$\mu$m diameter disks. The resist mask (SPR-955) was reflowed on a hotplate at 180~$^\circ$C and was used as an etch mask in an Ar/BCl$_3$/Cl$_2$ reactive ion etch (RIE) chemistry. By varying the optical exposure time it was found that rings could also be produced due to diffraction around the disk mask.

\begin{figure}
 \centering
  \centerline{\scalebox{0.5}{\includegraphics{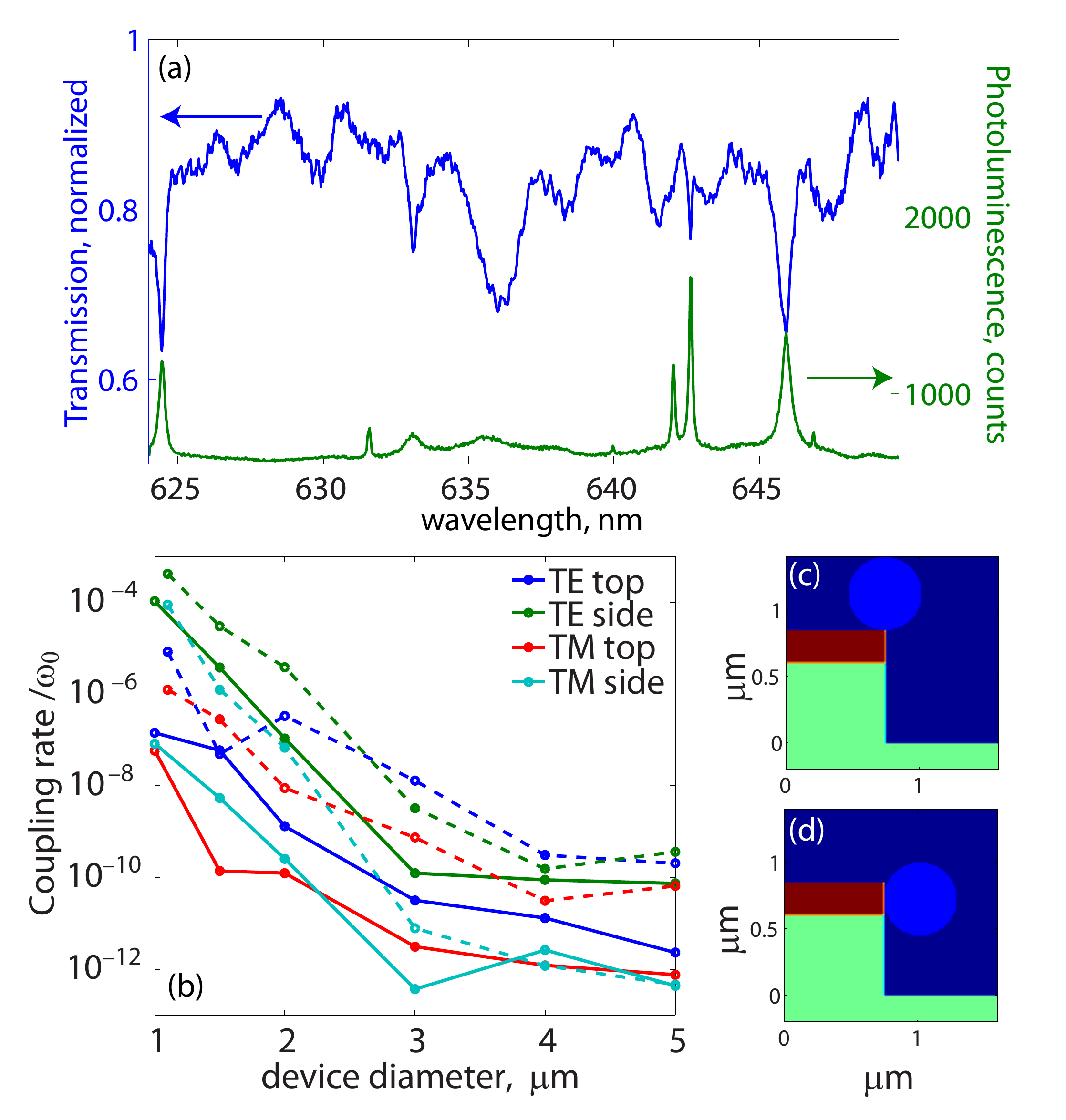}}}
  \caption{(a)Tapered-fiber photoluminescence and transmission spectra for device D1. Transmission dips indicating fiber-disk coupling coincide with PL peaks. (b) Estimation of the coupling rate for disk (solid lines) and ring (dashed lines) devices as a function of device diameter for the lowest order TE and TM modes near 637~nm. Two coupling geometries are calculated. In (c) the fiber is in contact with the top of the device. In (d) it is in contact with the side of the device. }
  \label{fig:taper}
\end{figure}

As a further step to facilitate tapered-fiber testing, a 20~$\mu$m tall, 190~$\mu$m wide diamond ridge defined by a SiN mask was etched on the diamond surface using an O$_2$/Ar RIE chemistry. The GaP microcavities were then transferred to the diamond ridge using the probabilistic ``drop-down'' method described in Ref.~\cite{ref:barclay2009cbm}. This procedure provides an easy shortcut for testing single devices, but eventually this must be replaced with a wafer-bonding approach before integrated photonic networks can be constructed. After transfer the devices were etched in an oxygen RIE plasma to create a 600~nm diamond pedestal beneath the GaP disks. This pedestal improves the lateral confinement of the cavity modes, decreasing radiation loss into the diamond substrate~\cite{ref:barclay2009hpc}.

Scanning electron microscope images of the two devices studied for this work are shown in Fig.~\ref{fig:devices}b,c. Device D1 consisted of a 250 nm thick, 1.5 micron diameter GaP disk and device D2 consisted of a 150~nm thick, 280 nm wide, ring waveguide with a 1.6~$\mu$m outer diameter. These devices were examples of the smallest devices made and had cavity resonances within tuning distance of the \NVm~ZPL.

\section{Tapered-fiber coupling to high-index cavities}

A 550~nm$\pm20$~nm diameter tapered fiber was used to collect NV photoluminescence from the hybrid GaP-diamond cavities. The tapered fiber position was controlled by a piezo stage and contacted to the top or side of the GaP microcavity to collect NV photoluminescence as shown in Fig.~\ref{fig:devices}a. Room-temperature transmission measurements were performed to characterize the fiber-microcavity coupling. Only in the smallest disk devices (sub-2~$\mu$m diameter) could transmission dips be observed as shown in Fig.~\ref{fig:taper}a. Transmission dips ranging from a few percent to a few tens of percent could be attained. The utility of using a tapered fiber for device evaluation depends critically on the collection efficiency, especially in the case of single centers. We have thus estimated the energy decay rate from the cavity mode into the waveguide as a function of device diameter numerically using coupled mode theory.

We first simulated the uncoupled lowest order TE and TM cavity modes for disk and ring devices using FDTD (MEEP~\cite{ref:oskooi2010mee}). The fiber mode was calculated using well-known analytic expressions for a step-index circular waveguide~\cite{ref:yariv1997oem}. The fiber diameter was taken to be 550~nm and the fiber index $n_f$ = 1.45. In Ref.~\cite{ref:manolatou1999cma}, an expression for the coupling rate between a cavity mode and fiber is derived using
power conservation. Using this formalism, one can show that the total decay rate of energy from the cavity into the fiber is given by
\begin{equation}\label{eq:cm}
\gamma_e = \left|\omega_0\int_{-l}^{l} dy \int_{A_f} dA (n_f^2-1) \vec{E}_f(\vec{r})\vec{E}_c^*(\vec{r})\right|^2
\end{equation}
where $\vec{E}_c$ and $\vec{E}_f$ are the cavity and fiber electric fields, respectively, and $\omega_0 = 2\pi c/(637\,\textrm{nm})$ is the resonant frequency. The 2D integral is taken over the fiber cross section $A_f$ and the 1D integral is along the fiber propagation axis $y$  with $\vec{E_c}\cong 0$ at $y = \pm l$. We have assumed that the cavity mode is a standing wave which decays equally into both forward and backward
propagating waveguide modes. The cavity field is normalized to unit energy~\cite{ref:kamalakis2006fdc},
$$
\mu\int_V\vec{H}_c^*(\vec{r})\cdot\vec H_c(\vec r) dV + \int_V \epsilon(\vec r)\vec E_c^*(\vec r) \cdot \vec E_c (\vec r) dV = 1
$$
and the waveguide field is normalized to unit power,
$$
\int_{A_f}(\vec E^*_f(\vec r)\times\vec H_f(\vec r) + \vec E_f(\vec r)\times \vec H_f^*(\vec r)) \cdot \hat y\,dA = \pm 1.
$$
$\vec{H_c}$($\vec{H_f}$)  is the cavity (fiber) magnetic field and $\epsilon$ is the dielectric constant.

The results of the coupled-mode calculation are shown in Fig.~\ref{fig:taper}b for both ring and disk devices for a fiber in contact with the top of the device~(Fig.~\ref{fig:taper}c) and for a fiber in contact with the side of the device~(Fig.~\ref{fig:taper}d). Experimentally there is uncertainty in the fiber-cavity contact position, the extent of the fiber-cavity contact, the fiber diameter, and the the phase of the standing-waves. Even with these uncertainties the simulations do provide some insight into how to achieve the best fiber-device coupling. As the device diameter becomes smaller it appears easiest to couple to the TE modes from the side of the disk. The ring device in general provides better coupling due to the thinner GaP layer. If we assume a reasonable cavity Q of 10,000, appreciable coupling (decay rate due to the taper is greater than 10\% of the uncoupled decay rate) only occurs for device diameters less than 1.5~$\mu$m. To be able to attain critical coupling in these geometries the device diameter would need to be sub-micron. Such small devices are desirable for large Purcell factors and could be fabricated using electron-beam lithography. There is however a fundamental limit on the minimum device diameter $d_\mathrm{min}$ before Q is limited by radiation loss, $Q_\mathrm{rad}(d_\mathrm{min})\cong 10,000$. FTDT calculations show that for GaP-diamond hybrid devices this diameter can be quite small: for a 150~nm GaP layer $d_\mathrm{min}\cong 700$~nm and for a 250~nm layer $d_\mathrm{min}\cong 650$~nm.

\section{Fiber collection of the cavity-coupled zero-phonon emission at 10K}

\begin{figure}
 \centering
  \centerline{\scalebox{1}{\includegraphics{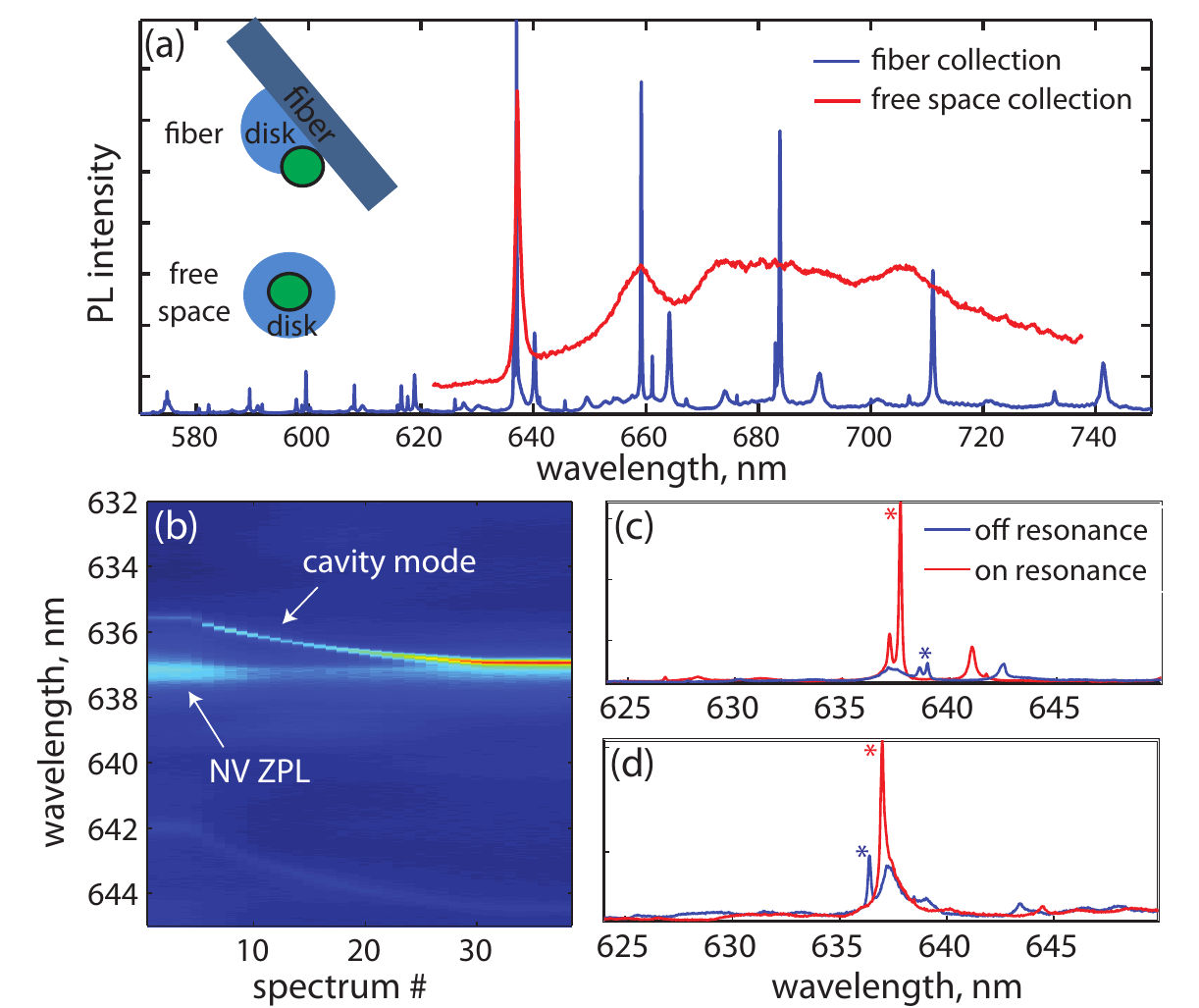}}}
  \caption{(a) Comparison of the NV spectra with free-space collection and with tapered-fiber collection for device D1. (b) Tuning spectra for device D2. Cavity-mode tuning with the introduction of Xenon gas can be observed. (c,d) NV spectra collected through the tapered fiber for D1,D2 with the cavity tuned on and off the zero-phonon resonance. The cavity mode of interest is labeled with an asterisk.}
  \label{fig:tuning}
\end{figure}

NV photoluminescence spectra collected in free space from the center of disk D1 and collected through a tapered fiber coupled to the whispering-gallery modes of the cavity are shown in Fig.~\ref{fig:tuning}a. In all PL measurements the NV centers are excited with 532~nm excitation. In the free space spectrum, excitation and PL collection is from the top of the disk through a 0.55 numerical aperture objective. The \NVm~ZPL line is observed at 637 nm with a 1.5~nm inhomogeneous linewidth. This inhomogeneity is generally observed in heavily implanted samples and could be a result of strain introduced during the implantation process. In the tapered fiber spectrum the disk is excited at its edge and the excitation spot is spatially separated from the tapered-fiber contact position. This is to reduce scatter from uncoupled NVs into the tapered fiber. The sharp cavity modes can clearly be observed over the background PL with a contrast greatly exceeding what can be achieved in free-space using spatial filtering~\cite{ref:barclay2009cbm}.

Due to variations in device fabrication, it is unlikely that a cavity mode will initially be on resonance with the 637nm zero-phonon emission. Temperature tuning is not useful in this system because the ZPL becomes phonon broadened at temperatures greater than $\sim$15~K~\cite{ref:fu2009odj}. Instead we tune the cavity by injecting xenon gas into the cryostat which condenses on the microcavity and alters its mode volume~\cite{ref:srinivasan2007oft}. Microdisk resonance tuning up to approximately 3~nm is possible for these GaP devices which have a free spectral range exceeding 20~nm. It is thus necessary to fabricate many devices to ensure a subset of devices is within the tuning range. Fig.~\ref{fig:tuning}b depicts a series of PL spectra from device D2 as Xe is continuously leaked into the cryostat. An increase in the PL intensity as the cavity is tuned onto the ZPL resonance is observed. A comparison between the on-resonance and off-resonance fiber spectra for both D1 and D2 are shown in Figs.~\ref{fig:tuning}c,d showing a marked enhancement in the ZPL collection in the resonant case. Quality factors of 4900 and 3500 were measured for D1 and D2 respectively.

\section{Lifetime measurements of cavity-coupled NV centers}

\begin{figure}
 \centering
  \centerline{\scalebox{1}{\includegraphics{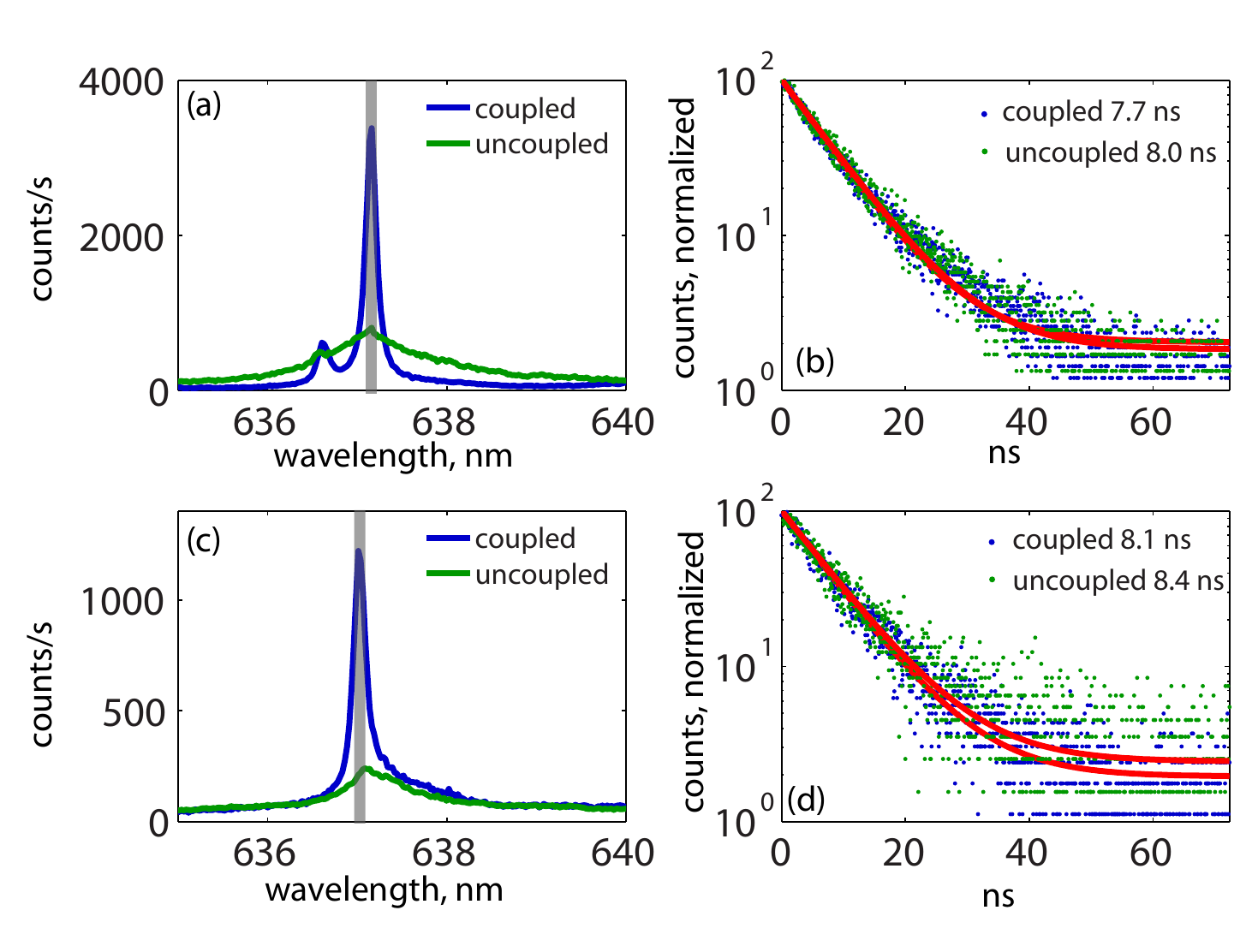}}}
  \caption{(a,c) PL spectra of coupled and uncoupled NVs in device D1, D2. Filter bandwidth of the collected PL for the lifetime measurement is indicated in grey. (b,d) Lifetime measurements of coupled and uncoupled NVs in device D1,D2.}
  \label{fig:lifetimes}
\end{figure}

The NV-cavity coupling strength can be determined from changes in the spontaneous emission lifetime that occur when the NV ZPL is resonant with the cavity mode. The decay rate for a single NV center prepared in a particular excited state $i$ may be written as,
\begin{equation}
\gamma_i = \gamma_0 + F_{p, i} \gamma_{0, \mathrm{ZPL}} \, ,
\label{eq:decayrate}
\end{equation}
where $\gamma_0$ is the total uncoupled decay rate including nonradiative channels as well as spontaneous emission through other modes, $\gamma_{0,\mathrm{ZPL}}$ is the zero-phonon spontaneous emission rate in bulk diamond, and $F_{p, i}$ is the Purcell factor. In high-quality diamond samples typically $\gamma_0 \approx (12 \, \mathrm{ns})^{-1}$, but in the present samples we observe $\gamma_0 \approx (8 \, \mathrm{ns})^{-1}$, suggesting a strong non-radiative component. We have previously measured $\gamma_{0, \mathrm{ZPL}} \sim 0.03$ of the total \NVm~photoluminescence intensity including the phonon sidebands which is consistent with the literature~\cite{ref:davies1974vsd}. If $\gamma_{0, \mathrm{ZPL}}$ represents a fraction $\delta$ of the total uncoupled decay rate $\gamma_0$ then the ratio of the on-resonance lifetime $\tau_c$ to the uncoupled lifetime $\tau_0$ would be
$$
\frac{\tau_c}{\tau_0} = \frac{\gamma_0}{\gamma_0+ F_{p,i}\gamma_\mathrm{ZPL}} = \frac{1}{1+F_{p,i}\delta}.
$$

 Using the tapered-fiber collection, we measure the SE decay curve of the cavity-coupled NV ensemble. This measurement, combined with a theoretical consideration of the effects of ensemble averaging, can be used to estimate the spontaneous emission enhancement of a single NV center with a particular position and orientation, as is discussed below. Prior to performing ensemble lifetime measurements the cavity was tuned to the ZPL resonance. NV emission was collected through the fiber using the geometry depicted in Fig.~\ref{fig:tuning}a. PL spectra under continuous-wave excitation of NVs in both the disk D1 and ring D2 devices are shown in Fig.~\ref{fig:lifetimes}a,c. For the lifetime measurements, a green pulsed excitation source was used, based on supercontinuum generation with a Ti-Sapphire laser. The excitation source had a bandwidth of 28~nm centered at 520nm and a repetition rate of 4.75~MHz. The ZPL emission was filtered as depicted by the shaded regions in Fig.~\ref{fig:lifetimes}a,c. The SE decay curves for devices D1 and D2 were fitted to a single exponential with decay times of $7.66\pm0.08$~ns and $8.08\pm0.18$~ns respectively (Fig.~\ref{fig:lifetimes}b,d). The measurement was repeated on NV centers that were not coupled to the cavity mode. To collect uncoupled emission the fiber was placed across the center of the device and the NV centers directly under the fiber were excited. As seen from the PL spectra in Fig.~\ref{fig:lifetimes}a,c, in this excitation/collection geometry the ZPL line is weak and broad and thus shows no features of the cavity mode. Lifetime measurements of uncoupled NVs yielded characteristic decay times of $7.95\pm0.08$~ns and $8.38\pm0.18$~ns for D1 and D2 respectively. We show below that this lifetime reduction is somewhat smaller than what is expected theoretically.

\section{Simulated lifetime modification}

Next, let us discuss the lifetime modification expected theoretically for a large ensemble of NV centers coupled to the cavity modes of device D2. Each NV center has a number of characteristics that affect its excited-state dynamics, including its position relative to the cavity mode, the orientation of the N-V axis along one of the four $\langle 111 \rangle$ axes, the principle axes of symmetry breaking for the two excited orbital states as determined by local strain or electric fields, and the frequency shifts of the optical transitions, again determined by local fields. Nonradiative decay channels may also be present, including decay through the singlet states (especially from the $m_s = \pm 1$ spin sublevels~\cite{ref:batalov2008tcp}).

To simplify the discussion, here we neglect the dynamics of the spin sublevels, and consider only the two orbital excited states. Such an approximation is not suitable for a detailed analysis of NV dynamics, but suffices to illustrate the effect of spatial and orientational inhomogeneity on the radiative dynamics. Under this simplification we can use Eq.~\ref{eq:decayrate} to model the decay rate for a single NV center prepared in a particular excited state $i$, and we let $\gamma_{0, \mathrm{ZPL}} = (0.03)(12\,\textrm{ns})^{-1} = 0.0025\,\textrm{ns}^{-1}$. The Purcell factor is given by,
\begin{equation}
F_{p, i} \approx \frac{3}{4\pi^2} Q \frac{(\lambda_0/n_\mathrm{c})^3}{V}
\frac{n_\mathrm{c}}{n_d} \left| \frac{\hat{\mu_i} \cdot \vec{E}(\vec{r})}{E_\mathrm{max}}
\right|^2 \frac{\kappa^2}{\kappa^2 + 4\Delta_i^2} \, ,
\end{equation}
which is similar to Eq.~\ref{eq:Purcell_simple1} except that we have included the factor of $n_c/n_d$ necessary for the hybrid geometry and also the effect of dipole alignment and spectral detuning. Here, $n_d$ is the diamond refractive index and $\hat{\mu_i}$ is a unit vector parallel to the relevant transition dipole moment. The hybrid-cavity mode volume is given by
$$
 V = \frac{\int d^3\vec{r}\epsilon(\vec{r})|\vec{E}(\vec{r})|^2}{\epsilon(\vec r_\mathrm{max})|\vec E (\vec r_\mathrm{max})|^2},
 $$
in which $\epsilon(\vec r_\mathrm{max}) = n_c^2$ is the dielectric constant at the location of the field maximum~\cite{ref:santori2010nqo}. The NV center has two optical transitions involving the two orbital excited states with dipole moments orthogonal to the $\langle 111 \rangle$ N-V axes. The principle axes~\cite{ref:hughes1967uss,ref:davies1976oso} can be described in terms of an angle $\beta$; for example for an NV oriented along $[1,1,1]/\sqrt{3}$, the dipole moments are $\mu_1 = \vec{u_1} \cos\beta + \vec{u_2} \sin\beta$ and $\mu_2 = -\vec{u_1} \sin\beta + \vec{u_2} \cos\beta$, where $\vec{u}_1 = [-1,-1,2]/\sqrt{6}$ and $\vec{u}_2 = [1,-1,0]/\sqrt{2}$. If the angle $\beta$ is determined by local fields then we expect it to be randomly distributed, and here we assume a uniform distribution. It is important to note that we neglect population relaxation between orbital excited states. This assumption is reasonable only for measurements performed at liquid-helium temperatures~\cite{ref:fu2009odj}. We further assume that the two orbital excited states have the same unmodified decay rate. In general, we must also include the effect of detuning, and here $\Delta_i$ is the detuning between the dipole transition frequency and the central frequency of the cavity mode, while $\kappa = \omega_0 / Q$ is the cavity linewidth. In the measurement, we detect photoluminescence through a spectral filter of $0.06 \, \mathrm{nm}$ bandwidth centered at the cavity frequency, and therefore we detect emission only from a subset of NV transitions that are nearly on resonance with the cavity mode. Thus we set $\Delta_i = 0$ for the present calculation.

\begin{figure}
 \centering
  \centerline{\scalebox{0.7}{\includegraphics{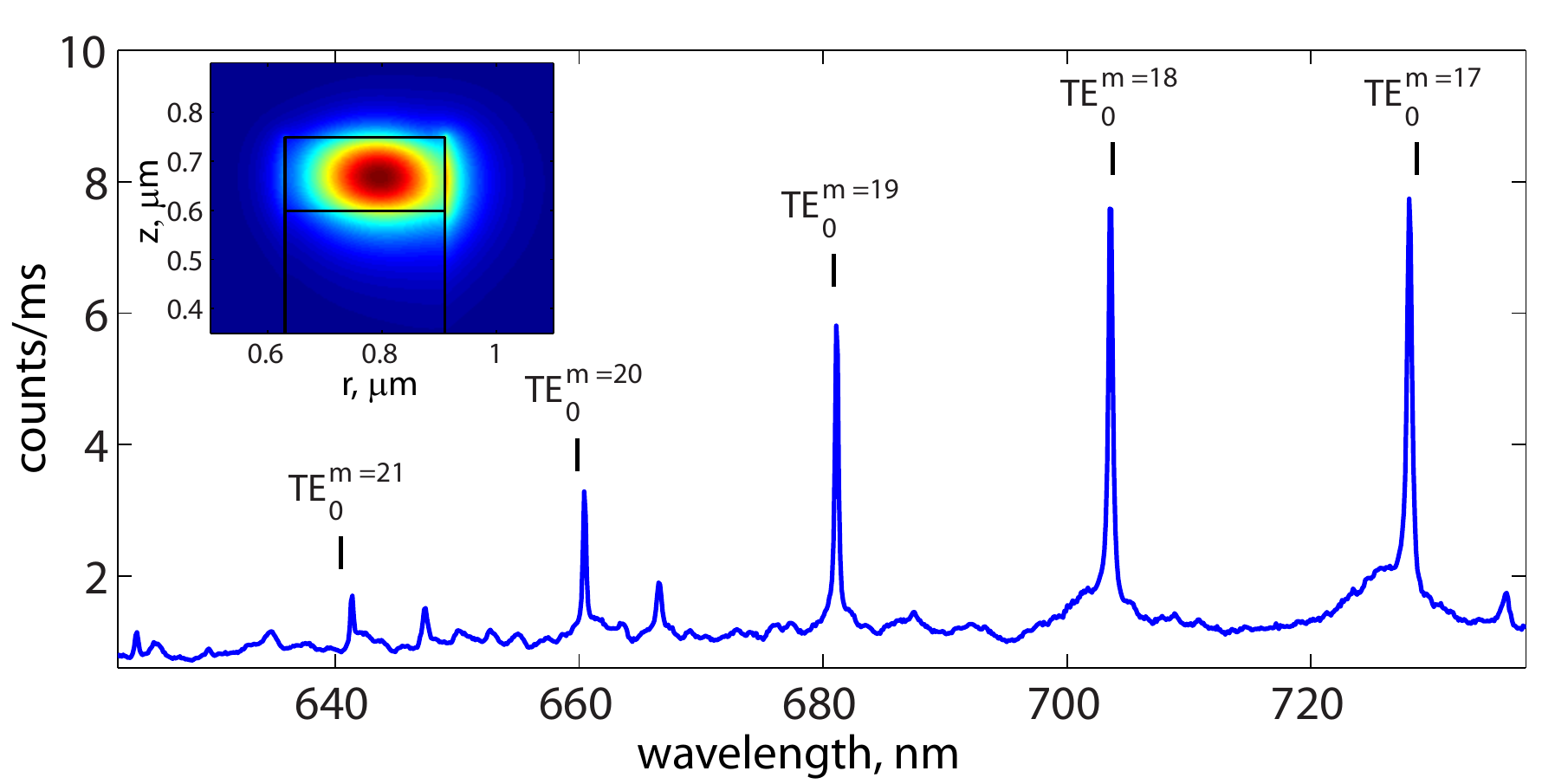}}}
  \caption{Room-temperature PL spectra collected through the tapered fiber for D2. Observed resonances are in reasonable agreement with a simulated modes which include the effect of dispersion in the GaP~\cite{ref:pikhtin1977rlg}. A similar agreement can be found for TM$_0$ modes for a ring with an outer diameter of 1.96~$\mu$m. \emph{Inset}: Simulated mode profile $|E(r,z,\phi=0)|$ for TE$_0^{m=21}$.}
  \label{fig:sim}
\end{figure}

While our model assumes a simple exponential decay for each NV transition, the ensemble-averaged behavior is non-exponential, but can be written as,
\begin{equation}
I(t>0) = \int_0^\infty d\zeta \, G(\zeta) e^{-(\gamma_0 + \zeta \gamma_{0, \mathrm{ZPL}}) t} \, ,
\end{equation}
where $I(t)$ is the measured PL decay curve, and $G(\zeta)$ represents a distribution of NV transitions with various Purcell factors, decaying at various rates, weighted by their relative occurrence and by their excitation and collection efficiencies. For free-space collection this distribution can be written as,
\begin{equation}
G_\mathrm{f.s.}(\zeta) \propto \sum_{\vec{r}, \vec{\alpha}, \beta} \rho(\vec{r},\vec{\alpha},\beta) I_\mathrm{exc}(\vec{r},\vec{\alpha},\beta) \, \delta(F_p(\vec{r},\vec{\alpha},\beta)-\zeta) \, .
\label{eq:G_freespace}
\end{equation}
Here, $\vec{\alpha}$ is the N-V orientation vector, and $\rho(\vec{r},\vec{\alpha},\beta)$ is the NV density with respect to position and dipole orientation. For the spatial dependence, we approximate $\rho(z)$ as a Gaussian distribution with respect to depth below the surface, with a mean depth of $15 \, \mathrm{nm}$ and variance (straggle) of $5 \, \mathrm{nm}$. Such a distribution corresponds to a uniform $10 \, \mathrm{keV}$ nitrogen ion implantation as estimated using SRIM~\cite{ref:zeigler2008sri}, assuming that the resulting NV centers incorporate only implanted nitrogen, and that the implanted nitrogen does not diffuse during annealing. The excitation efficiency $I_\mathrm{exc}(\vec{r},\vec{\alpha},\beta)$ includes a spatial dependence that is modeled as a diffraction-limited Airy disk pattern corresponding to $532\,\mathrm{nm}$ excitation through a 0.55 numerical-aperture lens ($\mathrm{FWHM} \sim 500 \, \mathrm{nm}$), centered between the inner and outer edges of the ring structure. The excitation efficiency is also assumed to be proportional to $|\vec{\mu} \cdot \vec{E}_\mathrm{exc}|^2$, where the excitation polarization $\vec{E}$ is along $[1,0,0]$.

\begin{figure}
 \centering
  \centerline{\scalebox{1}{\includegraphics{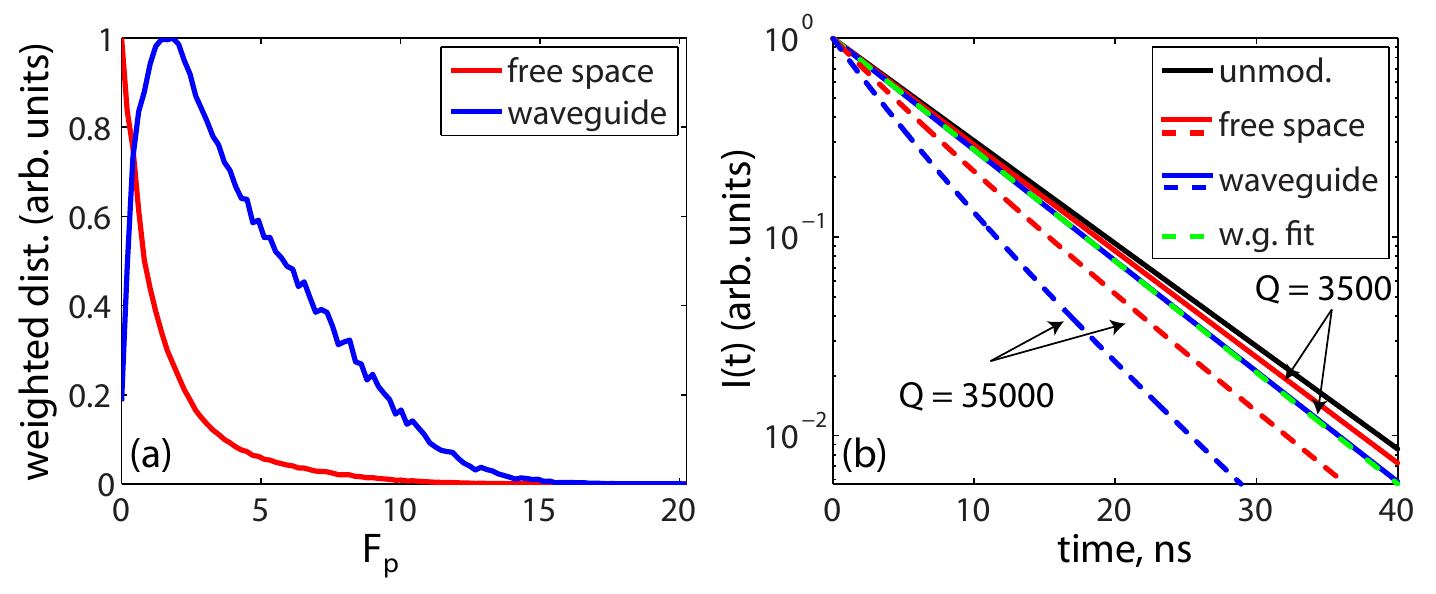}}}
  \caption{(a) Weighted distribution of the Purcell factor $F_p$ simulated for the D2 microring structure with free-space collection (red) and for collection through a waveguide coupled only to the cavity mode of interest (blue). (b) Simulated photoluminescence decay curves for unmodified NV centers with $\gamma_0 = (8.4\,\mathrm{ns})^{-1}$ (see text), for NV centers in the D2 microring structure detected with free-space collection (red), and for waveguide collection (blue). A single-exponential fit is also shown for the waveguide collection case (green, dashed) with $\tau=7.74$~ns. Simulated curves for devices with Q = 35000 are also included for comparison.}
  \label{fig:Fp_dist}
\end{figure}

For collection through the tapered fiber, Eq.\ref{eq:G_freespace} must be modified since, ideally, only light emitted into the cavity mode of interest is collected. Thus we weight by the emission rate into the cavity mode, which is proportional to the Purcell factor:
\begin{equation}
G_\mathrm{w.g.}(\zeta) \propto \sum_{\vec{r}, \vec{\alpha}, \beta} \rho(\vec{r},\vec{\alpha},\beta) I_\mathrm{exc}(\vec{r},\vec{\alpha},\beta) F_p(\vec{r},\vec{\alpha},\beta) \, \delta(F_p(\vec{r},\vec{\alpha},\beta)-\zeta) \, .
\label{eq:G_waveguide}
\end{equation}

The lowest-radial order TE cavity field was simulated for the calculation using FDTD (MEEP) software~\cite{ref:oskooi2010mee}. A reasonably good agreement between the experimental and simulated resonance frequencies could be obtained for a model device with a 1.8~$\mu$m outer diameter as shown in Fig.~\ref{fig:sim}. Using the parameters given above, setting $Q=3500$ and considering a single standing-wave cavity mode (rather than degenerate counter-propagating modes) we obtain the Purcell-factor histograms shown in Fig.~\ref{fig:Fp_dist}a. For this mode the maximum expected Purcell factor for an NV 15~nm below the surface and with perfect dipole alignment is $F_{p, \mathrm{max}} = 13$. However, for free-space collection the weighted average is only $\langle F_p \rangle = 1.8$. For collection through the tapered fiber, because we preferentially detect NV centers that are better coupled to the cavity mode, the average increases to $\langle F_p \rangle = 4.2$. But even with the tapered fiber it is clear that the random spatial and angular distributions of dipole moments reduces the mean cavity enhancement by a large amount.

For a more direct comparison with experiment, PL decay curves generated from these distributions are shown in Fig.~\ref{fig:Fp_dist}b. Here we have assumed $\gamma_0 = (8.4 \, \mathrm{ns})^{-1}$ to match the behavior observed in these heavily implanted samples, and $\gamma_{0, \mathrm{ZPL}} = 0.0025 \, \mathrm{ns}^{-1}$, as discussed above. Although these decay curves do not follow a single-exponential time dependence, because the average decay modification is weak compared with $\gamma_0$, it would be quite difficult experimentally to detect the deviation from single-exponential behavior caused by variations in $F_p$. The effective decay lifetimes obtained by fitting the first $40\,\mathrm{ns}$ of decay to a single exponential are $8.10 \,\mathrm{ns}$ and $7.74\,\mathrm{ns}$ for free-space and tapered-fiber collection, respectively. These are approximately equal to those expected from the $\langle F_p \rangle$ values given above: $8.09 \,\mathrm{ns}$ and $7.72 \,\mathrm{ns}$ for free-space and tapered-fiber collection, respectively. Fig.~\ref{fig:Fp_dist}b also includes simulations for a device with Q = 35000. Q's exceeding 25000 have been observed previously in much larger hybrid GaP-diamond devices~\cite{ref:barclay2009cbm} and more work is required in determining the factors limiting the Q in the present devices.

The lifetime modification shown in Fig.~\ref{fig:lifetimes}, which was at the limit of detection, is smaller than that predicted in this simulation, and some likely causes include the presence of an air gap between the GaP ring~\cite{ref:fu2008cnv} and the diamond surface, or non-ideal NV distributions. The simulations show that collection using the tapered fiber provides an important advantage since it selectively filters out NV centers that are poorly coupled to the cavity. Nevertheless, even using the tapered fiber for collection, ensemble averaging reduces the effective Purcell factor by a factor of approximately 3 compared with that expected for an ideally positioned and oriented NV center.

\section{Outlook}

There is significant motivation to integrate photonics with quantum systems for quantum communication, quantum information interfaces, and for quantum information processing. In final devices, quantum-photonic "nodes" will consist of single emitters, however single emitters are challenging to work with when studying and developing the classical photonic networks. Here we presented a method combining an ensemble of emitters and tapered-fiber collection which can be used for fast serial testing of a large number of QIP devices.

Experimentally we showed that tapered-fiber collection provides a way to obtain high-contrast spectra to reliably observe cavity modes. This is because the fiber enables the selective detection of NVs coupled to the cavity. Compared to a free-space collection, the measurement more directly probes the signal that would be collected by an on-chip waveguide, as part of an integrated photonic network. Although not exploited in this work, variable coupling is also possible with the tapered fiber providing a way to optimize the combined cooperativity parameter that includes both $F_p$ and the waveguide-cavity coupling efficiency. Additionally, numerical calculations indicate that for sub-micron GaP-diamond hybrid devices, the cavity energy decay rate into the fiber should be able to exceed the uncoupled decay rate allowing for efficient photon collection.

The use of NV ensembles provides two main advantages over single centers for device testing. First, the collected photoluminescence can be much greater allowing for faster testing. Second, device characterization does not rely on nanometer-scale placement of single emitters. Although the devices used to test this platform did not exhibit a significant Purcell enhancement, theoretically we showed that the fiber-coupled average Purcell factor is 2-3 times greater than that of free-space collection. Moreover, if the distribution of emitters is known, the ensemble device performance can also be used to estimate the single-emitter device performance.

This material is based upon work supported by the Defense Advanced Research Projects Agency under Award No. HR0011-09-1-0006 and The Regents of the University of California.

\section*{References}

%\bibliographystyle{iopart-num}
%\bibliography{lsip_qit,fu_njp2011}
\providecommand{\newblock}{}

\end{document}